# Use of extended and prepared reference objects in experimental Fourier transform X-ray holography

H. He[1], M. Howells[2], S. Marchesini[3], H. N. Chapman[4], U. Weierstall[1], H. A. Padmore[2] and  J.C.H.Spence[1]

1 *Department of Physics and Astronomy, Arizona State University, Tempe, Arizona 85287-1504*
2 *Advanced Light Source, Lawrence Berkeley Laboratory, 1 Cyclotron Road, Berkeley, California 94720*
3 *Lawrence Livermore National Laboratory, 7000 East Avenue, Livermore, California 94550*

Abstract

The use of one or more gold nanoballs as reference objects for Fourier Transform holography (FTH) is analysed using experimental soft X-ray diffraction from objects consisting of separated clusters of these balls. The holograms are deconvoluted against ball reference objects to invert to images, in combination with a Wiener filter to control noise. A resolution of ~30nm, smaller than one ball, is obtained even if a large cluster of balls is used as the reference, giving the best resolution yet obtained by X-ray FTH. Methods of dealing with missing data due to a beamstop are discussed. Practical prepared objects which satisfy the FTH condition are suggested, and methods of forming them described.

Lensless X-ray FTH (XFTH) provides a simple geometry for magnified imaging which avoids the need for a beam-splitter and the demanding coherence requirements of an off-axis reference beam[1], while also providing a simple solution to the phase problem for non-crystalline objects. Since only one Fourier transform is required with digital reconstruction, FTH is far superior over all iterative phase retrieval methods in terms of computation time. The method allows for complex objects, and is also capable of three-dimensional (tomographic) reconstruction if the unknown object and the reference object both lie on the rotation axis. FTH may be applied to many types of radiation – here we consider soft X-ray imaging, which allows diffraction data to be collected from biological cells. Despite these appealing attributes, the development of FTH has been slow. The best resolution obtained since the first introduction of FTH by Stroke[2,3] and Winthrop and Worthington[4] in 1965  was ~ 60nm[5] , as demonstrated by McNulty et. al. [6] in 1992 using zone plate lens under irradiation of coherent undulator soft X-rays.

By comparison, the resolution of other forms of X-ray holography have entered the atomic resolution stage [7-9]. Two difficulties may account for the slow development of high resolution XFTH. The first concerns the contradictory requirements on the size of the reference object. For high resolution, one normally requires a reference much smaller than the sample. However, for weakly scattering spheres of radius $s$ larger than the wavelength, the scattered wave intensity is proportional to $s^4$. The reference-wave contribution therefore will usually be much weaker than that of the sample. This often renders a high contrast hologram and high quality reconstruction impractical. The second factor concerns the difficulty of fabricating samples with the correct geometry. An object and a reference-scatterer, lying in the same plane normal to a coherent beam, must be arranged with a separation larger than the sum of the two objects' widths (the FTH condition). This is difficult to arrange for sub-micron objects. These difficulties have been the main barrier preventing FTH from becoming a practical imaging method rather than a laboratory demonstration. In this paper, we aim to address the first problem and make suggestions for the second. We show with experimental data that high resolution can be obtained even when using a reference object as large as the sample, if suitable deconvolution procedures and reference objects are used. We also propose solutions to the sample geometry problem using modern lithographic techniques.

Whereas the possibility of deconvoluting FTH for improved resolution has been discussed from the beginning [2], a practical deconvolution method for X-ray imaging which prevents noise amplification was first discussed in detail by Howells[10]. Howells also discusses the special class of objects (with flat scattering distributions) which allow high resolution reconstruction after deconvolution. The devonvolution discussed here is a procedure to retrieve the existing information hidden by the blurring of the reference object. It is essential that the scattering of the objects under study, especially the reference object, extend over the whole detector for highest possible resolution. For a detailed discussion of the trade-off between resolution improvement and noise amplification, see [11]. Howells et al[10] suggested that if a reference function (here denoted as $g$) or its spectrum ($G$) is known, one can deconvolute the hologram $I_H$ against $G$, and then apply an inverse Fourier transform. To minimize the effect of zero-crossings in $G$, a Wiener filter[12] is used. Thus

$$\psi_r(x) = F^{-1}\left\{ I_H \frac{G}{|G|^2 + \Phi} \right\} \quad (1)$$

where $\psi_r$, the final reconstruction, contains the desired unknown object function, $\Phi$ is the ratio of the noise power to signal power and is usually treated as a constant and $F^{-1}$ denotes inverse Fourier transform. In this way, a high resolution reconstruction may be possible, depending on the form of $G$, and the extent to which noise is controlled by the filter. Desirable edges and discontinuities in the object function will ensure that $G$ extends to high frequencies compared to $I_H$ [11], thus reducing the likelihood of division by zero in the presence of noise. An experimentally convenient reference object consists of colloidal gold balls, which have reasonably sharp edges, are simple for modeling, and are readily available in known nanometer sizes. We have therefore implemented the above approach using these balls as the reference for deconvolution-enhanced FTH (DEFTH).

Below we present a reconstruction of an experimental soft X-ray hologram based on the above ideas. Data from sample labeled as Au50107[13] was used, which consists of clusters of 50 nm diameter colloidal gold balls lying on a transparent silicon nitride membrane. We refer the reader to our early work [13,14] for a detailed description of the experiment. Transmission X-ray diffraction patterns were recorded at 588 eV in the far-field using a CCD camera. The objects on the sample (clusters of gold balls) are well separated, which enabled us to undertake a DEFTH analysis.

Sample Au50107 consists of two clusters of almost identical size. One cluster, taken as the reference, was first imaged by both the hybrid input output (HIO) algorithm [15] and the Difference Map algorithm [16]. As will be discussed later, the modeling of the reference is less important than the accuracy of the algorithm. Both the above mentioned algorithms were found to work well for DEFTH analysis, and are superior to direct simulation of the reference from SEM pictures. Fig. 1 (a) shows the resulting reconstruction of the hologram using deconvolution against the known cluster (which had been imaged using the HIO algorithm), and a Wiener filter to control noise. To demonstrate the validity of the results, we show in fig. 1 (b) a scanning electron microscope (SEM) image of the same cluster together with the reference cluster. We note here that the Wiener filter[12] is crucial for successful deconvolution, without it the image is completely unrecognizable.

In fig. 1 (a), artifacts in the background and on the object are evident. Careful inspection of the data shows that it originates from spurious structures produced by missing data in the center of the hologram, due to the use of a beam stop. The discontinuity at the beam stop, when propagated into real space, produces spurious structure which interferes with the true image. One way to minimize this effect is to replace the missing data with a smooth function such as a Gaussian. A better way appears to be the following. Starting from fig. 1 (a) and the reference object, one obtains a full image of the sample from which the power spectrum may be obtained. We then use this spectrum to fill in only the missing region of the hologram. Such a "renewed" hologram is then used for a second DEFTH reconstruction. Fig. 1 (d) shows the result of the above process repeated 10 times. As expected, the background "noise" is reduced significantly and the image of the balls becomes much smoother. A drawback of this procedure is the slightly increased computation time.

The most significant and surprising result here is the resolution achieved compared to the size of the reference. An investigation based on convoluting a simulated ball with a sinc function[17] suggests a reconstructed image resolution of 30 nm. (The resolution of the hologram extends to less than 10 nm). This is about 10 times smaller than the dimension of the reference cluster. The sharp drop at the edge of the ball ensures the presence of high frequency information in the hologram[11] and Wiener deconvolution is the key to retrieving this information. The problem of improving resolution has thus been transformed into one of accurately modeling the reference, rather than reducing the size of the reference. For the irregularly shaped object in sample Au50107 we found it is convenient to obtain the reference object using an independent imaging algorithm. This extra procedure does not create a problem for the applicability of DEFTH if the reference can be reused since only a single prior modeling is needed.

The last problem is how to arrange the sample and reference objects so as to satisfy the FTH condition, as discussed in the beginning of the paper. We propose the following method. A square lattice of heavy metal disks (or gold balls) with lattice parameter $a$ is deposited lithographically onto a silicon nitride membrane window. The lattice parameter should be large enough so that when a sample is placed in the middle of any neighboring two balls, one has *a/2 > (d + $d_d$)*, where *d* and $d_d$ are the diameter of the object and disks (or balls) respectively. The FTH condition can then be guaranteed no matter where the sample appears on the window. This is illustrated in fig 2. The X-ray beam (drawn as a gray area) should be of the correct diameter so as to cover just the object and its nearest *two* disks. In special cases (shown in the inset of fig. 2 enclosed by a dashed line), the object is close to the middle of the common line between neighboring disks, then the beam must cover only the object and *one* of its neighboring disks. This requirement can be dynamically monitored using the autocorrelation function of the scattering pattern during data collection. This ensures that there are no overlapping convolution objects in the autocorrelation pattern.

For a general unknown object (such as a cell), this work suggests that lensless imaging with X-rays should be possible at high resolution if a known object (such as one or more gold balls of known dimensions) can be placed beside it, and that significant resolution improvement can be obtained by deconvolution using this type of reference object in combination with a Weiner filter. Other deconvolution methods and filters exist which remain to be evaluated, while modern lithography methods may allow development of prepared reference objects which show even better noise/resolution performance under deconvolution. We are actively developing the use of metal deposition in a dual-beam focused-ion beam microscope for this purpose.

This work was supported by Department Of Energy under contract DE-AC03-76SF00098.

**Fig. 1. (a)** Reconstruction of a cluster of gold balls (sample Au50107) from soft X-ray transmission diffraction patterns using deconvolution and a Wiener filter. Constant $\Phi = 100$ in equation (1) produces best results. **(b)** SEM image of the same cluster (upper one), the reference (lower one) and their relative position. **(c)** Reconstruction using the iterative DEFTH algorithm (see text) to reduce the effect of missing data due to a beam stop. The diameter of each gold ball is about 50 nm.

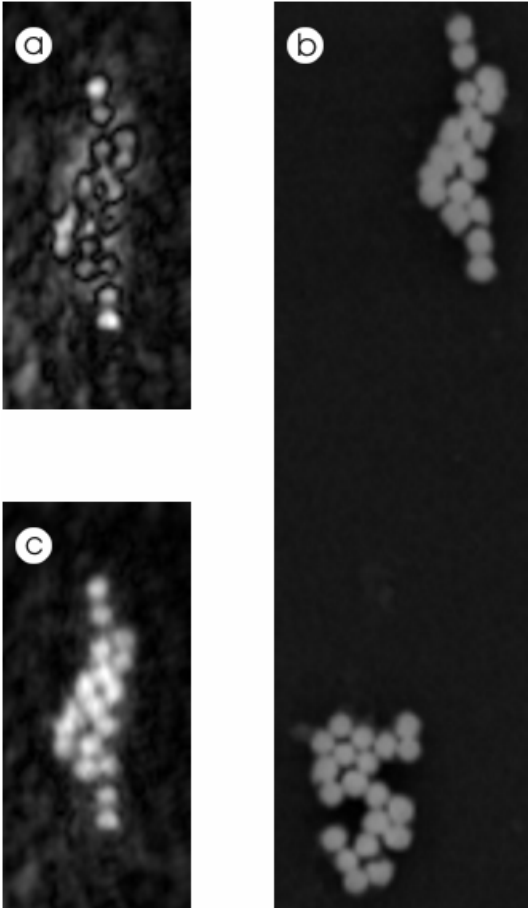

**Fig. 2.** Diagram of a sample substrate that always satisfies the FTH condition. Two typical cases are drawn to show how to determine the beam coverage area (gray color) depending on the relative positions between disk lattice (the reference) and the unknown object, one in the center of the figure and one at the bottom right corner of the figure (enclosed by dashed line). See text for details.

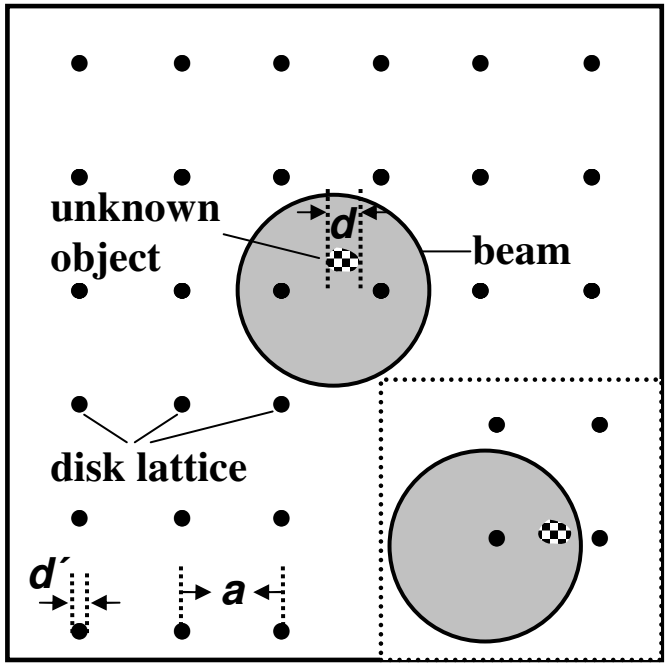

References:

[1]  R. J. Collier, C. B. Burckhardt, and L. H. Lin, *Optical Holography* (Academic Press, New York, 1971.).
[2]  G. W. Stroke and D. G. Falconer, Phys. Lett. **13,** 306-309 (1964).
[3]  G. W. Stroke, Appl. Phys. Lett. **6,** 201-203 (1965).
[4]  J. T. Winthrop and C. R. Worthing, Physics Letters **15,** 124 (1965).
[5]  The configuration used there has to rely on a "lens" (zone plate) to split the beam. Such a configuration is no better than using the same lens to build an X-ray microscope in terms of resolution.
[6]  I. McNulty, J. Kirz, C. Jacobsen, E. H. Anderson, M. R. Howells, and D. P. Kern, Science **256,** 1009-1012 (1992).
[7]  G. R. Harp, D. K. Saldin, and B. P. Tonner, Physical Review Letters **65,** 1012-1015 (1990).
[8]  M. Tegze and G. Faigel, Nature **380,** 49-51 (1996).
[9]  B. Sur, R. B. Rogge, R. P. Hammond, V. N. P. Anghel, and J. Katsaras, Nature **414,** 525-527 (2001).
[10] M. R. Howells, C. J. Jacobsen, S. Marchesini, S. Miller, J. C. H. Spence, and U. Weirstall, Nuclear Instruments & Methods in Physics Research Section a-Accelerators Spectrometers Detectors and Associated Equipment **467,** 864-867 (2001).
[11] P. E. Batson, D. W. Johnson, and J. C. H. Spence, Ultramicroscopy **41,** 137-145 (1992).
[12] K. R. Castleman, Prentice-Hall. 1979**,** xvi+429.
[13] H. He, S. Marchesini, M. Howells, U. Weierstall, H. Chapman, S. Hau-Riege, A. Noy, and J. C. H. Spence, Physical Review B-Condensed Matter **67,** 174114 (2003).
[14] H. He, S. Marchesini, M. Howells, U. Weierstall, G. Hembree, and J. C. H. Spence, Acta Crystallographica - Section a - Foundations of Crystallography **A59,** 143-52 (2003).
[15] J. R. Fienup, Journal of the Optical Society of America a-Optics Image Science and Vision **4,** 118-123 (1987).
[16] V. Elser, Journal of the Optical Society of America A-Optics & Image Science **20,** 40-55 (2003).
[17] Due to the fact that our imaging system cut off the high spatial frequency in a rectangular window, this corresponds to convolution of the image with a square sinc function in real space.